# Dominant role of $GABA_{B2}$ and Gßγ for $GABA_B$ receptor mediated-$ERK_{1/2}$/CREB pathway in cerebellar neurons

Haijun Tu[1,4], Philippe Rondard[2,3,4], Chanjuan Xu[1], Federica Bertaso[2], Fangli Cao[1], Xueying Zhang[1], Jean-Philippe Pin[2,3] and Jianfeng Liu[1,*]

[1] Sino-France Laboratory for Drug Screening, Key Laboratory of Molecular Biophysics (Huazhong University of Science and Technology), Ministry of Education, School of Life Science and Technology, Huazhong University of Science and Technology, Wuhan, Hubei, China.

[2] Department of Molecular Pharmacology, Institute of Functional Genomics, CNRS UMR5203, Université de Montpellier, F-34094 Montpellier, France;

[3] INSERM U661, F-34094 Montpellier, France;

[4] Co-first authors.

*Running title:* $GABA_{B2}$- and Gßγ-mediated ERK1/2 activation by $GABA_B$.

***To whom correspondence should be addressed :***

**Dr Jianfeng LIU**

Sino-France Laboratory for Drug Screening, Key Laboratory of Molecular Biophysics (Huazhong University of Science and Technology), Ministry of Education, School of Life Science and Technology, Huazhong University of Science and Technology, Wuhan, Hubei, China.
1037, Luoyu Road,
430074, Wuhan, Hubei, China.
Phone : +86 27 87792031
Fax : +86 27 87792024
Email : jfliu@mail.hust.edu.cn

**Abstract:**

γ- aminobutyric acid type B (GABA$_B$) receptor is an allosteric complex made of two subunits, GABA$_{B1}$ and GABA$_{B2}$. GABA$_{B2}$ plays a major role in the coupling to G-protein whereas GABA$_{B1}$ binds GABA. It has been shown that GABA$_B$ receptor activates ERK$_{1/2}$ in neurons of the central nervous system, but the molecular mechanisms underlying this event are poorly characterized. Here, we demonstrate that activation of GABA$_B$ receptor by either GABA or the selective agonist baclofen induces ERK$_{1/2}$ phosphorylation in cultured cerebellar granule neurons. We also show that CGP7930, a positive allosteric regulator specific of GABA$_{B2}$, alone can induce the phosphorylation of ERK$_{1/2}$. PTX, a G$_{i/o}$ inhibitor, abolishes both baclofen and CGP7930-mediated ERK$_{1/2}$ phosphorylation. Moreover, both baclofen and CGP7930 induce ERK-dependent CREB phosphorylation. Furthermore, by using LY294002, a PI-3 kinase inhibitor, and a C-term of GRK-2 that has been reported to sequester Gβγ subunits, we demonstrate the role of Gβγ in GABA$_B$ receptor mediated-ERK$_{1/2}$ phosphorylation. In conclusion, the activation of GABA$_B$ receptor leads to ERK$_{1/2}$ phosphorylation via the coupling of GABA$_{B2}$ to G$_{i/o}$ and by releasing Gβγ subunits which in turn induce the activation of CREB. These findings suggest a role of GABA$_B$ receptor in long-term change in the central nervous system.

1. Introduction

γ-aminobutyric acid (GABA) is a major inhibitory neurotransmitter in the central nervous system [1, 2]. It mediates fast synaptic inhibition through $GABA_A$ and $GABA_C$ ionotropic receptors as well as slow and prolonged synaptic inhibition through the metabotropic $GABA_B$ receptor [3]. $GABA_B$ receptor mediates both presynaptic inhibition of neurotransmitter release and post-synaptic inhibition of neuronal excitability [4, 5]. Accordingly, $GABA_B$ receptor is involved in various types of epilepsy, nociception and drug addiction. Activation of the receptor also has myorelaxant activity that are commonly used to treat spasticity associated with multiple sclerosis [2, 6]

$GABA_B$ receptor belongs to the class C of G-protein coupled receptors (GPCRs), together with metabotropic glutamate (mGlu), extracellular $Ca^{2+}$-sensing, and some pheromone and taste receptors [7]. Each of these receptors is composed of an extracellular domain called the Venus flytrap domain (VFT), to which agonists bind, and a heptahelical domain (HD), which is responsible for the recognition and activation of heterotrimeric G-proteins. Whereas mGluRs and $Ca^{2+}$-sensing receptors exist as homodimers, $GABA_B$ receptor is a heterodimer composed of two homologous subunits, $GABA_{B1}$ and $GABA_{B2}$ [8-11].

Heterodimerization of $GABA_B$ receptor is a prerequisite for $GABA_B$ receptor function. The VFT domain of $GABA_{B1}$ is sufficient for ligand binding [7], but its assembly with $GABA_{B2}$ increases the affinity of $GABA_{B1}$ for agonists [12]. Although $GABA_{B2}$ does not appear to bind any natural ligand [13], $GABA_{B1}$ needs to be associated with $GABA_{B2}$ to reach the cell surface [14]. Agonist binding to the $GABA_{B1}$ VFT domain results in its closure and this is likely responsible for a change in the relative position of the two VFT and HD domains [15]. This movement then allows activation of G proteins mediated by the HD domains of $GABA_{B2}$ [16, 17]. Recently, a novel class of compounds called $GABA_B$ positive allosteric modulators (PAM) such as CGP7930, appeared to represent a better therapeutic alternative than the classical agonist [18, 19]. Although CGP7930 alone was reported to be inactive in several assays, direct activation of the receptor by this positive allosteric modulator acting in the HD domain of $GABA_{B2}$ have also been observed [20].

Functional $GABA_B$ receptor is predominantly coupled to heterotrimeric $G_{i/o}$-type protein since most of $GABA_B$ receptor-mediated effects are inhibited by pertussis toxin (PTX) [21, 22]. Upon activation of the G protein, the $G\beta\gamma$ complex represses $Ca^{2+}$ influx by inhibiting $Ca^{2+}$ channels [23] and triggers $K^+$ channels opening [24, 25]. In the mean time, the $G\alpha_{i/o}$ subunits modulate the level of cyclic adenosine monophospate (cAMP) by regulating adenylate cyclase activities [26].

Protein phosphorylation plays a critical role in synaptic plasticity, learning and memory in vertebrates [27]. The Extracellular signal-Regulated protein Kinases 1/2 ($ERK_{1/2}$), also known as p42/44 mitogen-activated protein kinase (MAPK), signaling cascade plays important roles in the modulation of long-term potentiation in area CA1 of the hippocampus and is required for several forms of learning and memory [28]. Recently, it has been reported that $GABA_B$ receptor induces $ERK_{1/2}$ phosphorylation in the CA1 area of mouse hippocampus [29]. Regulation of $ERK_{1/2}$ signaling cascade by GPCRs is highly complex and cell type-specific [30], and the mechanism of $GABA_B$ receptor-mediated $ERK_{1/2}$ phosphorylation is still poorly understood. Furthermore, $GABA_B$ receptor is reported to bind to the transcription factor CREB2(cAMP responsive element binding protein-2)/ATF4(activating transcription factor 4) through coiled-coil interactions [31-33]. Nuclear translocation of CREB2 is also observed following $GABA_B$ receptor activation [31, 33]. The physiological significance for $GABA_B$ receptor activation-induced CREB2 translocation awaits elucidation.

In the current study, we examine the role of $GABA_B$ receptor in the regulation of $ERK_{1/2}$ and CREB phosphorylation in cultured cerebellar granule neurons. We find that selective $GABA_B$ receptor activation induces $ERK_{1/2}$ phosphorylation that in turn mediates CREB phosphorylation. We also show that this effect occurs via the GABA$_{B2}$ coupling to G$_{i/o}$ proteins by releasing Gβγ subunits. Interestingly, we show for the first time that selective activation of GABA$_{B2}$ is sufficient to induce CREB phosphorylation.

## 2. Materials and Methods

### 2.1 Materials

GABA was obtained from Sigma (Shanghai, China). Baclofen was purchased from Tocris (Fisher-Bioblock, Illkrich, France). GP54626, CGP7930, Pertussis toxin (PTX) were purchased from Calbiochem (US and Canada). Fetal bovine serum, culture medium, and other solutions used for cell culture were from Invitrogen (Shanghai, China). LY294002 and U0126 were purchased from Cell Signaling Technology (Beverly, MA). pRK5 plasmids encoding wild-type $GABA_{B1}$ and $GABA_{B2}$ with an epitope tag at their N-terminal ends under the control of a cytomegalovirus promotor were described previously [14]. The pcDNA3-c-myc-CD8-βARK plasmid, which was composed of the CD8 antigen membrane receptor and a domain containing the G-protein βγ subunit-binding site of GRK2, was a generous gift from Dr. M. De Waard.

### 2.2 Primary cerebellar granule neuronal cultures

Primary cultures of cerebellar cells were prepared as previously described [34]. Briefly, 1-week-old newborn mice were decapitated and cerebellum-dissected. The tissue was then gently triturated using fire-polished Pasteur pipettes, and the homogenate was centrifuged at 500rpm. The pellet was resuspended and plated in tissue culture dishes previously coated with poly-L-ornithine. Cells were maintained in a 1:1 mixture of DMEM and F-12 nutrient (Life Technologies, Gaithersburg, MD), supplemented with glucose (30mM), glutamine (2mM), sodium bicarbonate (3mM) and HEPES buffer (5mM), decomplemented fetal calf serum (10%), and 25mM KCl to improve neuronal survival. Three-days-old cultures contained $1.25 \times 10^5$ cells/cm$^2$.

### 2.3 Cell culture and transfection

Human embryonic kidney HEK293 were cultured in Dulbecco's modified Eagle's medium supplemented with 10% fetal bovine serum and transfected by electroporation as described previously [35]. Cells (10 x10$^6$) were transfected with plasmid DNA containing the cDNAs encoding GABA$_{B1}$ (4μg) or GABA$_{B2}$ (4μg), and completed to a total amount of 10μg of plasmid DNA with the pRK5 empty vector.

**2.4 Western blot analysis**

Cell lysates from cultures were sonicated, and protein concentrations were determined using the Bradford reagent (Bio-Rad Laboratories LTD, Hertfordshire, UK). Equal amounts of protein (20μg) were resolved by SDS-PAGE on 12% gels. Proteins were transferred to polyvinylidene fluoride membranes (Millipore, Bedford, MA) and blocked in blocking buffer (5% nonfat dry milk in TBS and 0.1% Tween 20) for 1 hr. The blots were then incubated with the primary rabbit polyclonal antibodies against phospho-ERK$_{1/2}$ (1:1000; Cell Signaling Technology, Beverly, MA), or with a rabbit polyclonal antibodies against the total ERK$_{1/2}$ (1:1000; Cell Signaling Technology), overnight at 4°C. This was followed by 1 hr incubation with goat anti-rabbit horseradish peroxidase (HRP)-linked secondary antibodies (1:20000; Cell Signaling Technology). Immunoblots were revealed using the enhanced chemiluminescence reagents (Pierce, USA) and visualized using the X-Ray film. The density of immunoreactive bands was measured using NIH image software, and all bands were normalized to percentages of control values.

**2.5 Drug treatment**

Cultures were washed once with HBS (Ca$^{2+}$ free) and preincubated at 37°C with HBS for 60 min. Cells were treated by adding freshly made drugs to the HBS. At the end of the treatment, cells were washed quickly with ice-cold PBS, pH 7.4 (Ca$^{2+}$ free), lysed with 200μL lysis buffer and placed immediately on ice. The cell monolayer was rapidly scraped in ice-cold lysis buffer. Drugs were dissolved in HBS with or without dimethyl sulfoxide (DMSO) or/and alcohol. Whenever DMSO or/and alcohol were used, HBS containing the same concentration of DMSO or/and alcohol were used as the control vehicle.

## 3. Results

### 3.1. Activation of GABA$_B$ receptor increases ERK$_{1/2}$ phosphorylation in neurons

We first studied the effect of GABA and the GABA$_B$-selective agonist baclofen, on ERK$_{1/2}$ phosphorylation in cultured mouse cerebellar granule neurons (CGNs). We found that GABA at 100μM caused a rapid and transient increase in ERK$_{1/2}$ phosphorylation with no changes in ERK$_{1/2}$ expression levels (Fig. 1A, upper panel). ERK$_{1/2}$ phosphorylation peaked at 10min and then decreased. Similar results were obtained with baclofen at 100μM (Fig. 1A, lower panel). These data showed that the activation of GABA$_B$ receptor leads to increased ERK$_{1/2}$ phosphorylation in CGNs.

To demonstrate that ERK$_{1/2}$ phosphorylation occurs via the specific activation of GABA$_B$ receptor, we evaluated the effect of the GABA$_B$ receptor selective antagonist on baclofen-induced ERK$_{1/2}$ phosphorylation in CGNs. Cells were pretreated for 20min with GABA$_B$ receptors antagonist, CGP54626 (10μM) and then stimulated with baclofen (100μM) for 10min. We found that this antagonist blocked baclofen-induced ERK$_{1/2}$ phosphorylation without to alter ERK$_{1/2}$ expression levels (Fig. 1B), thus demonstrating that GABA$_B$ receptor activation contributes to the baclofen-induced ERK$_{1/2}$ phosphorylation in CGNs.

### 3.2 GABA$_B$ receptor-mediated ERK$_{1/2}$ phosphorylation occurs through the coupling of GABA$_{B2}$ to G$_{i/o}$ protein

GABA$_B$ receptor, reported as a G$_{i/o}$ protein-coupled receptor activates its downstream effectors through coupling of GABA$_{B2}$ subunits towards G proteins [22, 23]. Recently, it was demonstrated that CGP7930, a positive allosteric modulator (PAM) of the GABA$_B$ receptor not only modulates GABA$_B$ receptor activity by directly acting on GABA$_{B2}$ HD domain but also activates GABA$_{B2}$ when expressed alone [36]. To evaluate whether GABA$_B$ receptor-mediated ERK$_{1/2}$ effect occurs through GABA$_{B2}$ coupling to G protein, we then studied the effect of CGP7930 on ERK$_{1/2}$ phosphorylation in CGNs. CGP7930 at a concentration of 50μM caused a rapid, transient and strong increase in ERK$_{1/2}$ phosphorylation in CGNs (Fig. 2A). ERK$_{1/2}$ phosphorylation levels induced upon CGP7930 treatment peaked at 10 min prior to decreasing gradually to the basal level. Pretreatment of CGNs with CGP54626 even at the concentration of 100μM did not abolish CGP7930-induced ERK$_{1/2}$ phosphorylation (Fig. 2A, inset panel). These results demonstrate that GABA$_{B2}$ directly activates ERK$_{1/2}$ pathway. In addition, the required association of GABA$_{B1}$ with GABA$_{B2}$ to reach the cell surface [14] indicates that GABA$_B$ receptor mediated- ERK$_{1/2}$ phosphorylation in CGNs occurs through GABA$_{B2}$ coupling to G protein.

We further verified that GABA$_B$ receptor mediated-ERK$_{1/2}$ phosphorylation is also regulated through coupling of G$_{i/o}$ protein by using pertussis toxin (PTX). Neurons were pretreated with PTX at 200ng/ml for 14-18hrs, and then stimulated for 10 min with 100μM baclofen or 50μM CGP7930 (Fig. 2B). Under those circumstances, PTX inhibited baclofen or CGP7930 -induced ERK$_{1/2}$ phosphorylation in CGNs (Fig. 2B), demonstrating that baclofen or CGP7930 induced ERK$_{1/2}$ activation via G$_{i/o}$ proteins.

We confirmed the observed GABA$_B$ receptor-mediated ERK$_{1/2}$ phosphorylation in HEK293 cells transfected with both GABA$_{B1}$ and GABA$_{B2}$, or with GABA$_{B2}$ alone. GABA at 100μM caused a rapid and transient ERK$_{1/2}$ phosphorylation without change in ERK$_{1/2}$ levels in cells expressing both GABA$_{B1}$ and GABA$_{B2}$ whereas it had no effect on the mock-transfected cells (cells transfected with pRK5 empty vector) (Fig. 3A). CGP7930 alone also induced ERK$_{1/2}$ phosphorylation (Fig. 3B), and this effect was not antagonized by a CGP54626 pre-treatment of the cells (Fig. 3B, inset panel). Pretreatment of the cells with PTX blocked both GABA and CGP7930-induced ERK$_{1/2}$ phosphorylation (Fig. 3B). These results demonstrate that GABA$_B$ receptor-mediated ERK$_{1/2}$ phosphorylation occurs via GABA$_{B2}$ coupling to G$_{i/o}$ protein. In HEK293 cells expressing GABA$_{B2}$ alone, CGP7930 induced an acute phosphorylation of ERK$_{1/2}$ whereas it had no effect on mock-transfected cells (Fig.

4A). Pretreatment of the cells with PTX efficiently reduced CGP7930-mediated ERK$_{1/2}$ phosphorylation with no change in ERK$_{1/2}$ expression levels (Fig. 4B), thus suggesting that the activation of GABA$_{B2}$ by CGP7930 is sufficient to mediate ERK$_{1/2}$ phosphorylation via G$_{i/o}$ protein.

Our results in CGNs and in HEK293 cells transfected with GABA$_{B2}$ alone or with both GABA$_{B1}$ and GABA$_{B2,}$ showed that GABA$_B$ receptor induced ERK$_{1/2}$ phosphorylation via the coupling of GABA$_{B2}$ to G$_{i/o}$ heterotrimeric proteins.

**3.3 Gβγ subunits mediate GABA$_B$ receptor-induced ERK$_{1/2}$ phosphorylation**

To test whether the coupling of GABA$_{B2}$ to G$_{i/o}$ protein occurs via Gβγ/PI-3 kinase to induce ERK$_{1/2}$ phosphorylation, we treated CGNs with the selective PI-3 kinase inhibitor, LY294002. Interestingly, the inhibition of the PI-3 kinase pathway led to a strong inhibition of both baclofen and CGP7930- induced ERK$_{1/2}$ phosphorylation in neurons (Fig. 5A), thus suggesting an important role for Gβγ derived from G$_{i/o}$ proteins.

We then investigated the role of Gβγ subunits in GABA$_B$ receptor-mediated ERK activation in HEK293 cells expressing the GABA$_B$ receptor. To this end, we used the previously characterized Gβγ-scavenger consisting of the C-terminal region of GRK2 (βARK) fused to the extracellular and transmembrane domains of CD8 which then provides a membrane anchor for βARK's C-tail (CD8-βARK)[37] [38]. The overexpression of CD8-βARK inhibited GABA-induced ERK$_{1/2}$ phosphorylation in HEK293 cells co-expressing both GABA$_{B1}$ and GABA$_{B2}$ and CGP7930-induced ERK$_{1/2}$ phosphorylation in HEK293 cells expressing GABA$_{B2}$ alone (Fig. 5B). Taken together, these results show that GABA$_B$ receptor mediates ERK$_{1/2}$ phosphorylation through Gβγ subunits.

**3.4 GABA$_B$ receptor-mediated ERK$_{1/2}$ phosphorylation induces CREB phosphorylation**

Phosphorylation of $ERK_{1/2}$ plays an important role in the regulation of gene expression via phosphorylation of nuclear transcription factors [39]. We therefore tested the response of CREB to the $GABA_B$ receptor mediated-$ERK_{1/2}$ pathway. We found that baclofen at 100μM induced a rapid and transient increase in CREB phosphorylation in CGNs (Fig. 6A). Baclofen-induced CREB phosphorylation was abolished by the pretreatment of CGNs with either CGP54626 (Fig. 6B) or $MEK_{1/2}$ inhibitor U0126 (Fig. 6C). Furthermore, the pretreatment of CGNs with U0126 also inhibited CGP7930-induced CREB phosphorylation (Fig. 6C). These results show that $GABA_B$ receptor-mediated CREB phosphorylation occurs through the $ERK_{1/2}$ pathway.

## 4. Discussion

The main findings of the present study concern the mechanism of $GABA_B$ receptor-mediated $ERK_{1/2}$ phosphorylation. We show that: 1) $GABA_B$ receptor activation leads to increased $ERK_{1/2}$ phosphorylation in cultured cerebellar granule neurons which in turn induces CREB phosphorylation; 2) selective activation of $GABA_{B2}$ by CGP7930 is sufficient for $ERK_{1/2}$ activation in both cultured cerebellar neurons and HEK293 cells transfected with $GABA_{B2}$ alone or in the presence of $GABA_{B1}$; 3) all these effects rely on a PTX-sensitive $G_{i/o}$ heterotrimeric protein-dependent pathway by releasing Gßγ and by implicating PI-3 kinase pathway, and 4) both baclofen and CGP7930-mediated CREB phosphorylation is $ERK_{1/2}$ dependent.

Several reports have recently shown that $G_{i/o}$-coupled $GABA_B$ receptor induced $ERK_{1/2}$ phosphorylation [29, 40-43]. However, the signaling cascades transmitting $GABA_B$ receptor signals towards $ERK_{1/2}$ remained unclear. Indeed, the biochemical routes linking GPCRs to $ERK_{1/2}$ are highly complex and cell type-specific [44]. Our results demonstrate the role of $Gα_{i/o}$ protein in selective activation of $GABA_B$ receptor to $ERK_{1/2}$ signaling. Furthermore, it is well established that PI3 kinase acts downstream of Gβγ subunits to mediate GPCR-controlled MAPK activation [45-47]. The inhibition of the PI-3 kinase leading to a strong inhibition of baclofen-induced $ERK_{1/2}$ phosphorylation in neurons suggests an important role for Gßγ derived from $G_{i/o}$ proteins for $GABA_B$ receptor-mediated $ERK_{1/2}$ activation.

We showed that $GABA_B$ receptor-mediated $ERK_{1/2}$ activation requires the presence of $GABA_{B2}$. These results are compatible with a recent study where $GABA_{B1}$ has been reported to reach the cell surface by co-expression with $GABA_A$ receptors γ2S subunits in heterologous cellular systems but failed to stimulate $ERK_{1/2}$ phosphorylation in the absence of $GABA_{B2}$ [43]. Furthermore, CGP7930, a positive modulator of $GABA_B$ receptor, was reported to potentiate $GABA_B$ receptor activity by interacting with HD domain of $GABA_{B2}$ [36]. However, the effect of CGP7930 alone on $GABA_B$ receptor activity is controversial. Meanwhile, CGP7930 produced little or no stimulation of the $GABA_B$ receptor activity in some studies [18], others show that CGP7930 alone can stimulate inositol phosphate accumulation in HEK293 cells co-expressing the $GABA_B$ receptor and the chimeric G-protein Gqi9 [36]. Here, we observed CGP7930 alone displays an intrinsic agonist activity on $GABA_B$ receptor mediated-$ERK_{1/2}$ activation in both CGNs and HEK293 cells expressing $GABA_B$ receptor. Competitive antagonist of $GABA_B$ receptors such as CGP54626 that binds on $GABA_{B1}$ did not block the CGP7930-mediated $ERK_{1/2}$ phosphorylation, suggesting that $GABA_{B2}$ activation is sufficient to induce $ERK_{1/2}$ activation. CGP7930-mediated ERK1/2 phosphorylation in HEK-293 cells expressing $GABA_{B2}$ alone is compatible with the effect of CGP7930 on inositol phosphate accumulation in HEK-293 expressing $GABA_{B2}$ alone [36].

Several studies suggested that $GABA_B$ receptor plays an important role in memory processing. For example, mice lacking the $GABA_{B1}$ [25] or the $GABA_{B2}$ subunit [48] suffer from severe memory impairment. Increasing evidence has shown that $ERK_{1/2}$ plays an important role in long-term synaptic plasticity and memory through regulating protein synthesis and gene expression [49, 50]. Stimulation of CREB is also critical for long term potentiation [51] through either $ERK_{1/2}$ or p38 pathway [52, 53]. In our study, we demonstrate that CREB phosphorylation is induced by selective activation of $GABA_B$ receptor via an $ERK_{1/2}$-dependent pathway. These data suggest a role of the ERK/CREB pathway in $GABA_B$ receptor-mediated long-term synaptic plasticity and memory.

**Acknowledgements**

This work was supported by the National Natural Science Foundation of China (Grants No. 30530820, No.30470368) and Hi-Tech Research and Development Program of China (863 project) (Grants No. 2006AA02Z326). H. T was supported by the fund of outstanding thesis from Huazhong University of Science and Technology (D0611). We would like to thank Dr Eric Chevet and Dr Jian Zhu for helpful discussions.

**Figure Legends:**

**Figure 1. *Activation of GABA$_B$ receptor increases ERK$_{1/2}$ phosphorylation in cultured mice cerebellar granule neurons.*** *A, Upper panel*, time course of the endogenous ERK$_{1/2}$ phosphorylation after incubation of GABA (100μM). Data represent the mean ± SEM from at least five independent experiments. *Lower panel*, typical immunoblots used to quantify the phosphorylated ERK$_{1/2}$ (pERK$_{1/2}$). *B,* Effects of the antagonist CGP54626 on baclofen-induced ERK$_{1/2}$ phosphorylation. CGP54626 (10μM) was incubated for 30 min before and during treatment with baclofen (10min). pERK$_{1/2}$ was quantified as previously in Fig. 1A, and data are the mean ± SEM from three independent experiments.

**Figure 2. *GABA$_B$ receptor-mediated ERK$_{1/2}$ phosphorylation occurs through the coupling of GABA$_{B2}$ to G$_{i/o}$-protein.*** *A,* Effect of the CGP7930 (50μM), a PAM of GABA$_{B2}$ in the increase of ERK$_{1/2}$ phosphorylation in cultured mice CGNs. Data are the mean ± SEM from three independent experiments. *Inset*, effect of CGP54626 (100μM) on CGP7930-induced ERK$_{1/2}$ phosphorylation. *B,* Inhibitory effect of PTX on baclofen- and CGP7930-stimulated ERK$_{1/2}$ phosphorylation. Data are the mean ± SEM from three independent experiments

**Figure 3. *GABA$_B$ receptor-mediated ERK$_{1/2}$ phosphorylation in HEK293 cells co-expressing both GABA$_{B1}$ and GABA$_{B2}$.*** *A,* Effect of GABA (100μM) in the increase of transient ERK$_{1/2}$ phosphorylation. Data are the mean ± SEM from five independent experiments. *B,* Inhibitory effect of PTX on GABA- and CGP7930-stimulated ERK$_{1/2}$ phosphorylation. *Inset*, Effect of CGP54626 (100μM) on CGP7930-induced ERK$_{1/2}$ phosphorylation. Data represent the mean ± SEM from at least three independent experiments.

**Figure 4. *Activation of GABA$_{B2}$ increases ERK$_{1/2}$ phosphorylation in HEK293 cells expressing GABA$_{B2}$ alone.*** *A,* Effect of CGP7930 (50μM) in the increase of transient ERK$_{1/2}$ phosphorylation. *B,* Inhibitory effect of PTX on GABA- and CGP7930-

stimulated ERK$_{1/2}$ phosphorylation. Data are the mean ± SEM from at least four independent experiments.

**Figure 5.** *Gβγ subunit mediates GABA$_B$ receptor-induced ERK$_{1/2}$ phosphorylation.* *A,* Effect of LY294002 (20μM) on GABA$_B$ receptor-mediated ERK$_{1/2}$ phosphorylation induced by baclofen (100μM) or CGP7930 (50μM) in CGNs. *B,* Effect of over-expression of c-myc-tagged CD8-ßARK, Gßγ subunits inhibiting peptide, on GABA and CGP7930-induced ERK$_{1/2}$ phosphorylation in HEK293 cells expressing the heterodimer GABA$_B$ and GABA$_{B2}$ alone, respectively. Data are the mean ± SEM from at least four independent experiments.

**Figure 6.** *GABA$_B$ dependent CREB phosphorylation is mediated by ERK$_{1/2}$ phosphorylation*. *A,* The effect of baclofen (100μM) on CREB phosphorylation (pCREB) was determined by immunoblotting, and quantified as previously for pERK$_{1/2}$ in Figure 1A. *B,* Effect of CGP54626 (10μM) on baclofen-induced CREB phosphorylation. *C,* Effects of U0126 (10μM) on baclofen- or CGP7930- mediated ERK$_{1/2}$ and CREB phosphorylation in CGNs. Data are the mean ± SEM from at least four independent experiments.

# Figure 1

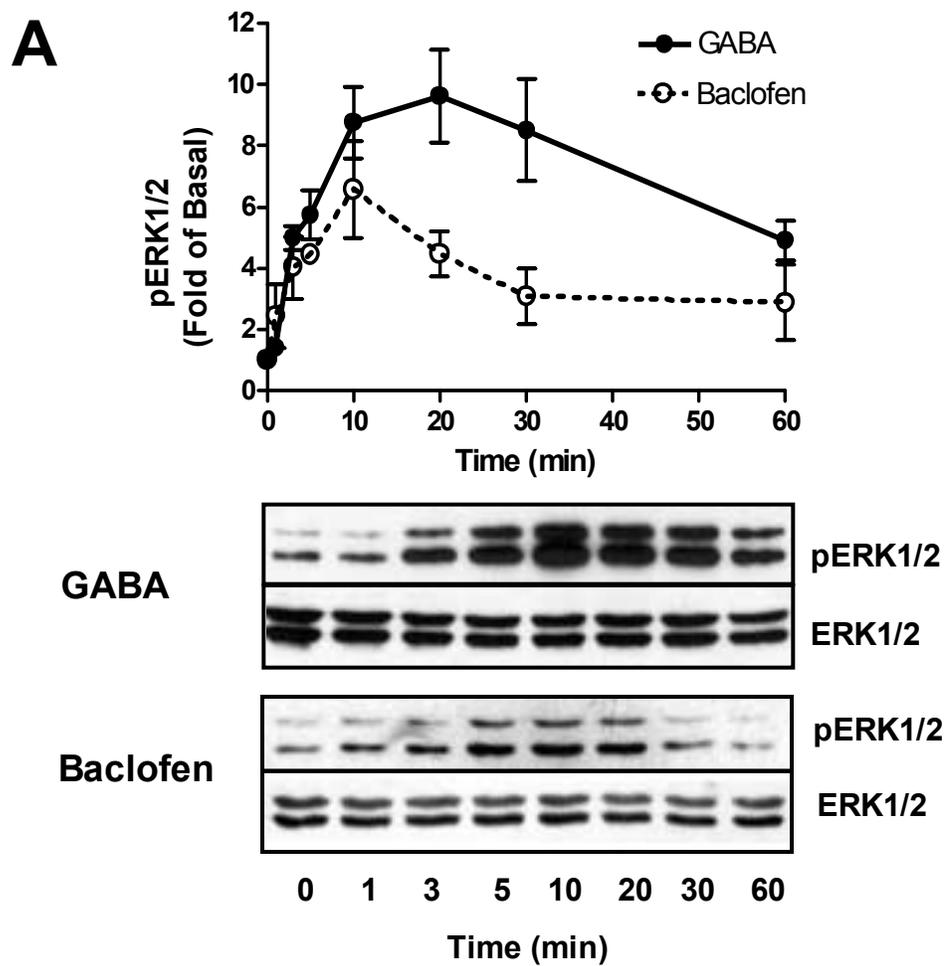

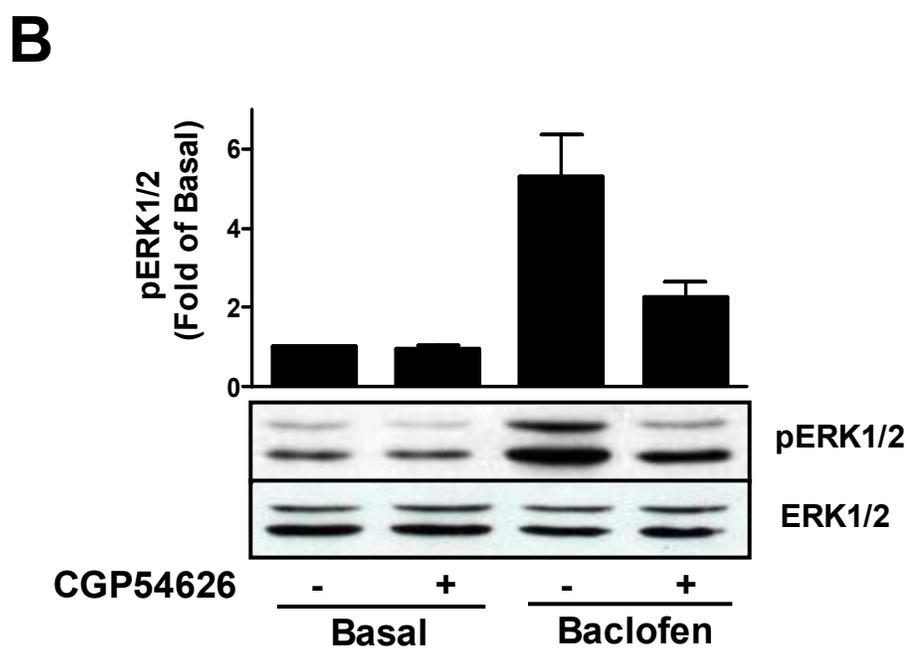

Figure 2

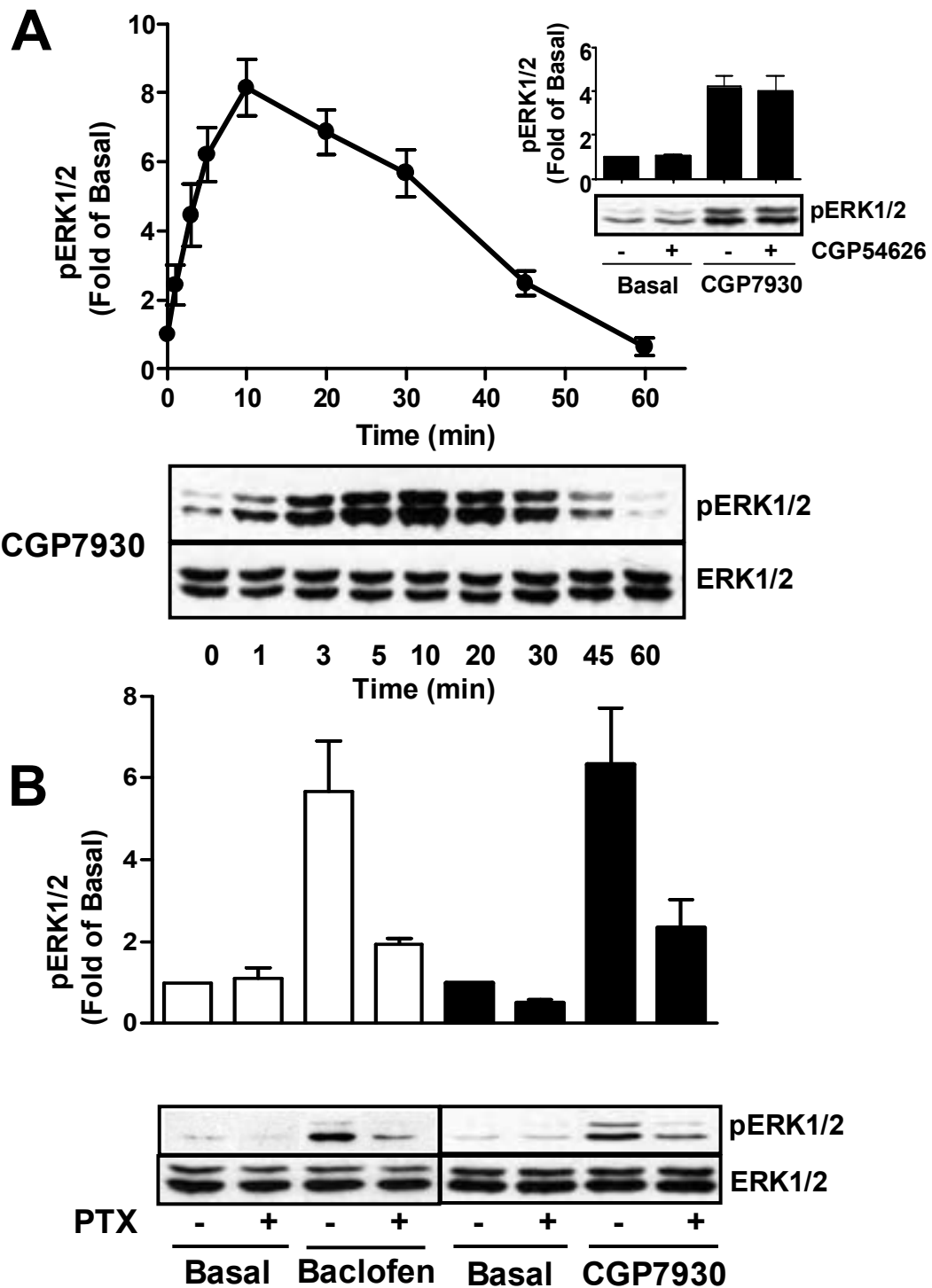

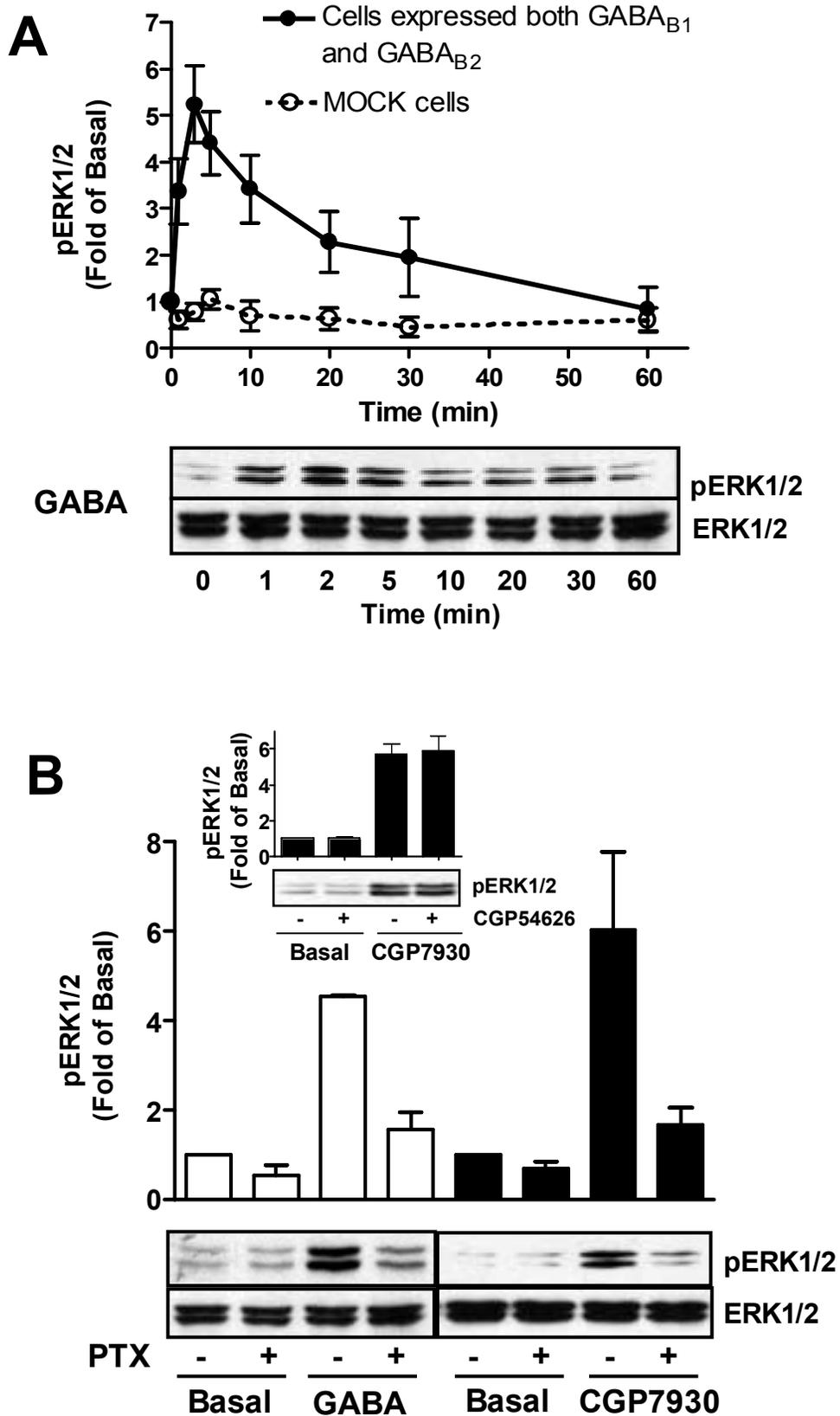

Figure 4

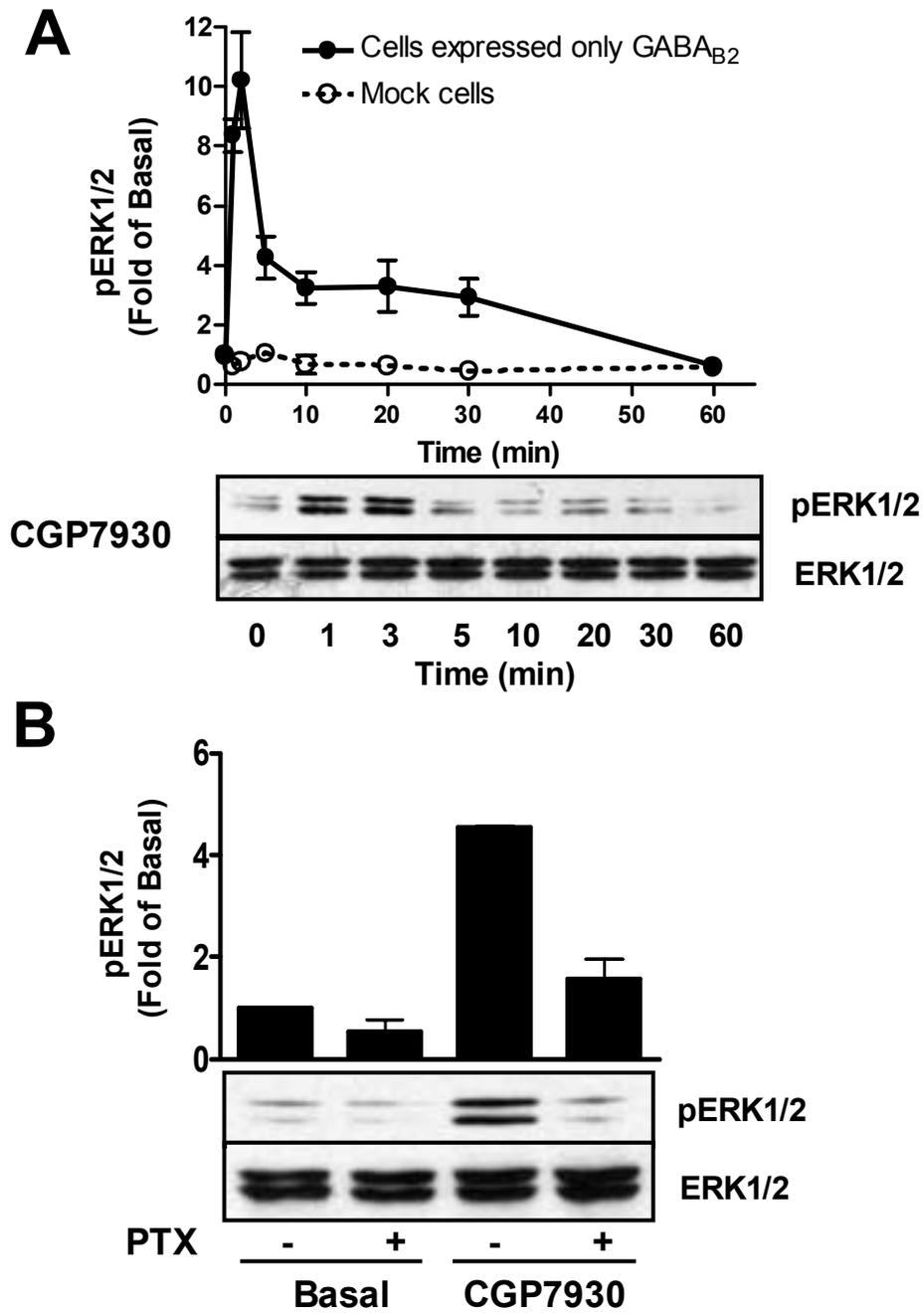

Figure 5

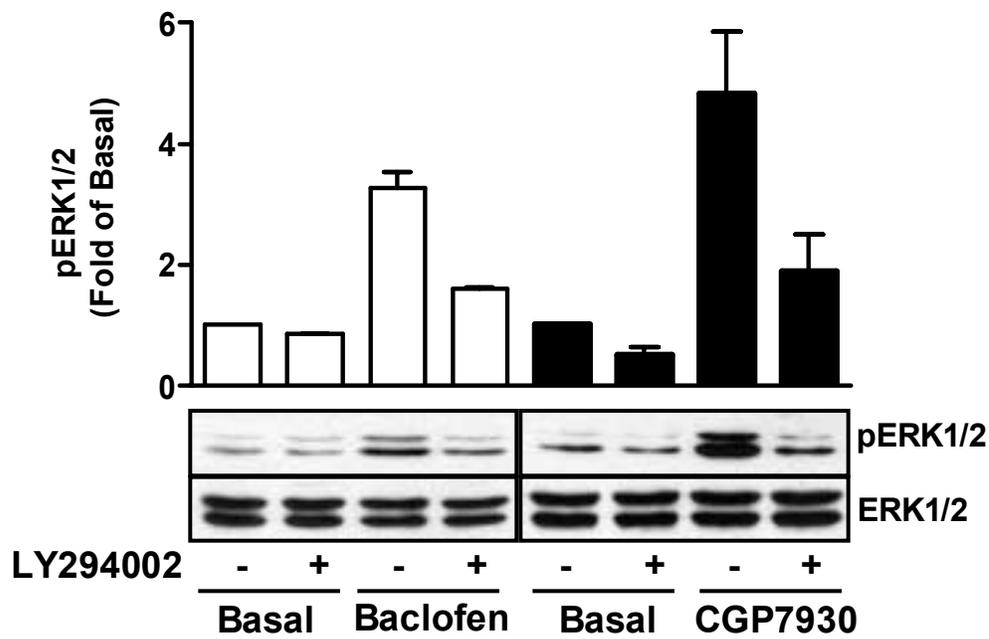

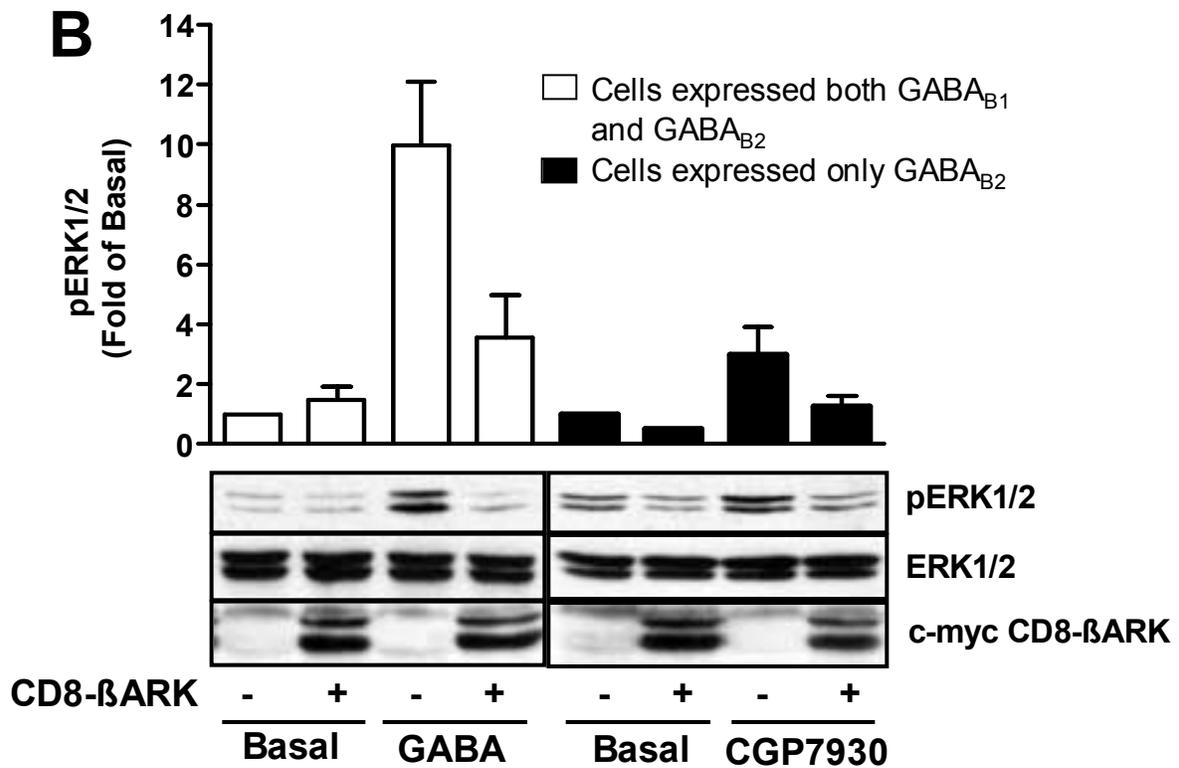

# Figure 6

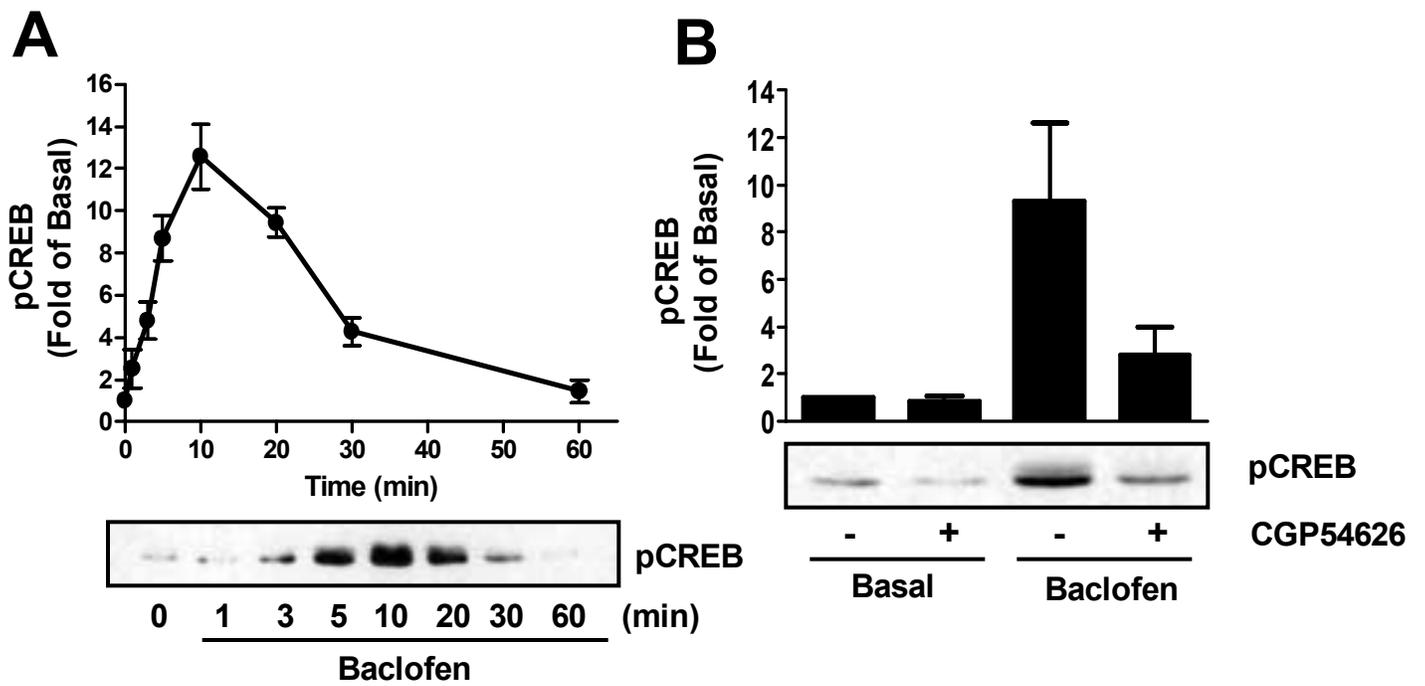

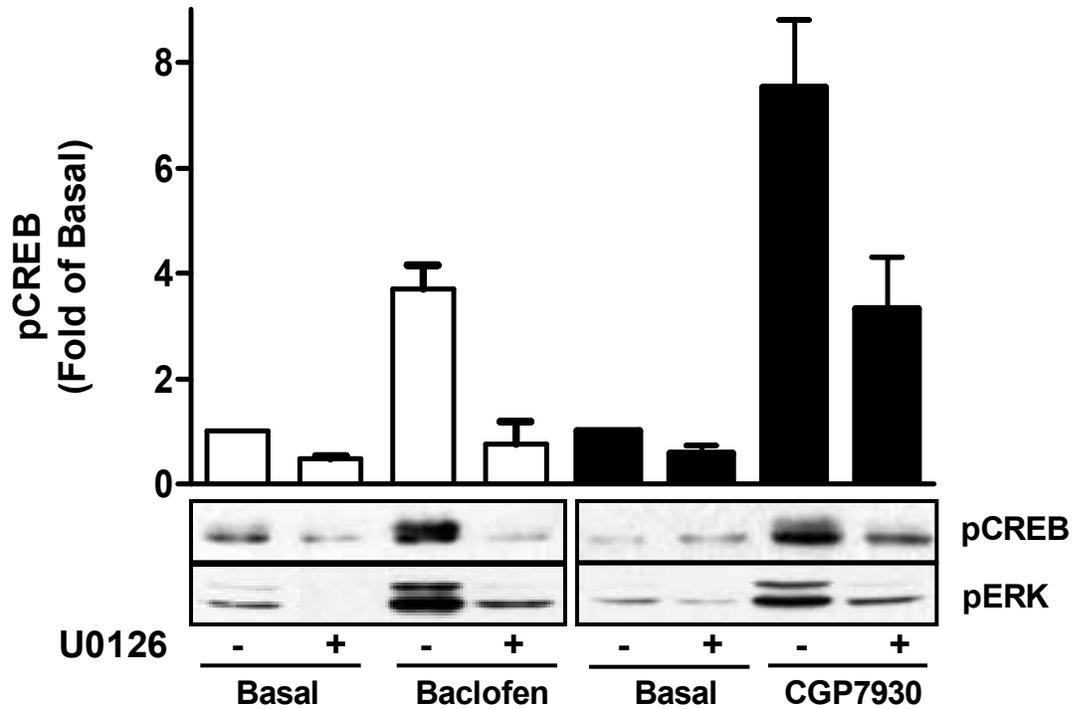